\documentclass[klunum]{kluwer}
\usepackage{epsfig}
\usepackage{rotating}
\usepackage{graphicx}
\usepackage{makeidx}
\makeindex

\begin{document}

\begin{article}
\begin{opening}
\title{Quasi-Periodicity in
	global solar radio flux at metric wavelengths during Noise Storms.}
\author{$ G. A. Shanmugha \surname{Sundaram}^{1, 2}$\email{sga@physics.iisc.ernet.in}}
\author{$ K. R. Subramanian^{2}$\email{subra@iiap.res.in}}
\runningauthor{Shanmugha Sundaram \& Subramanian}
\runningtitle{Quasi-Periodicity in Active-Sun radio flux}
\institute{}
\institute{1. Joint Astronomy Programme, Department of Physics,
        Indian Institute of Science, Bangalore - 560012, Karnataka, INDIA.}
\institute{2. Indian Institute of Astrophysics, Bangalore - 560034,
	Karnataka, INDIA.}
\date{March 24, 2004}

\begin{ao}
G. A. Shanmugha Sundaram\\
Joint Astronomy Programme\\
Department of Physics\\
Indian Institute of Science\\
Bangalore - 560012\\
Karnataka\\
INDIA
\end{ao}

\begin{abstract}
We present observational results from studying the
quasi-periodicities in global solar radio flux
during periods of enhanced noise storm activity,
over durations of $\sim4$~hrs a day ( "intra-day" variations ),
observed at 77.5 MHz with the newly commissioned
log-periodic array tracking system at the
Gauribidanur radio observatory.
Positional information on the
storm centers were obtained with the radio imaging data
from the Nan\c cay RadioHeliograph ( NRH ), while their
active region ( AR ) counterparts on the photosphere 
( and the overlying chromosphere ) were located
from the $H{\alpha}$ images of the Big Bear Solar Observatory ( BBSO ).
The quasi-periodicity in flux was found to be 110 minutes,
with the fluctuation in flux being $3 (\pm 1.5)$ solar flux
units ( sfu ). The results of such pulsations are interpreted
qualitatively as evidence for coronal seismology.

\end{abstract}


\end{opening}

\section{Introduction}
Type I solar noise storms, ever since their discovery in the
year 1946 \cite{hey46}, have proved to be amongst the most
profilic of events to occur at metric wavelengths.
They comprise of the bursts that are radio flux enhanced, narrowband
( df/f $\approxeq$ 3 \%  ), spiked ( 0.1 - 1 s ) events,
with the broadband ( df/f $\approxeq$~100\% ) continuum, lasting
from several hours to a few days, serving as their diffuse
background radio emission, on the dynamic spectral records.
According to \cite{hey46},~\cite{mcd47},
noise storms have their origins in the outer corona,
and appear proximal to the
sites of active regions in the photosphere and the chromosphere.\\
\\
In this paper, we present results from investigations on the
variations of global radio flux, during periods of prolonged
noise storm activity, with attributes to
quasi-periodicities on short time scales,
from observing the Sun continuously for about 6 hrs each day.
Details on the time-delay control enabled, broadband operable antenna
array system used for the study are provided, followed by a
note on the observation schedule adopted.
Interpretation of the imaging data from
complementary observations on the noise storms and the
associated underlying ARs constitute part of the latter section.
The scheme deployed to analyse and determine
the extent of global radio flux variations is described in
\S 4. The results of this study are presented in the Discussion
\& Conclusion section, along with qualitative remarks
on the plausible origin for the observed flux oscillations at metric
wavelengths, and their likely coronal implications.
\section{Antenna Instrumentation}
The Gauribidanur Radioheliograph ( GRH~\cite{ram98} ) is a transit-mode
instrument, while the time-delay control implemented on the
Gauribidanur Prototype Tracking System ( GPTS ) to GRH,
enables the Sun to be followed as it traverses the sky.
The antenna system comprises one group each along the
(E-W) and (N-S), as with the GRH, forming an L-shaped ($8 \times 4$)
element log-periodic dipole antenna ( LPDA ) array,
interconnected in a X-mas tree topology by radio frequency ( RF )
cables. The delay-tracking scheme has been implemented
on its front-end electronics, and involves a network of delay-line
cables and signal-loss compensating attenuators; this would eliminate
the otherwise cumbersome task of mechanically steering the
LPDA array, while also permitting radio
observations over a significantly wider frequency band.
The observing frequency was set at 77.5 MHz, and this choice
was based on the relatively terrestrial-interference-free
signals that were acquired over the entire observing duration.\\
\\
Automated PC-based control and monitoring of the tracking process
was deployed for an entire session of radio observations ( spanning
about 4 hours ), so as to achieve a precise "dwell-time" between adjacent
beam-positions, in accordance with a look-up table, while also serving
to implement the "$\delta$-setting", so as to attune the response of
the array to the declination ($\delta$)~of the source under study.\\
\\
The receiver system was of the analog superheterodyne type, with
a square law detector and a PC-based acquisition system constituting
the data detection, digitization, real-time display, and storage entities.
\section{Observations}
Continuous solar observations were made
at 77.5 MHz, over the period from 24th
June, 2002 to 20th August, 2002, with gaps in the observations at some
beam positions on a few days, assigned zero flux values subsequently. The daily
schedule involved solar observations at seven successive beam-positions,
comprising zenith-angles from $27^{0}$ E to $27^{0}$ W, in equiangular
steps of $9^{0}$, on either side of the local meridian at Gauribidanur,
for about four hours each day. A "sit-and-stare" method of acquiring the
solar profile was adopted, while the Sun drifted through the beam pre-deployed
at a particular position, as the in-phase and quadrature-phase components,
to yield the signal intensity as a convolved pattern of the Sun's disk
and the array power pattern.\\
\\
Absolute solar radio flux calibration
was performed, by following an identical observing schedule for
the intense, unresolvable radio source like Cygnus A, Cas A,
Crab, or Virgo, whose flux-densities are reliably known~\cite{bar77}.
This procedure, while translating the total power acquired by the receiver
( in arbitrary units of signal intensity ), for a given set of solar
observations, in terms of radio flux ( in janskys or sfu ), also ensures that
the gain variations, observed during tracking of the Sun at all
seven beam positions, are effectively equated at each such position.
The beam position dependent gain-variations are a result of ohmic losses
in the passive components on the delay-cards, frequency-dependent
phase characteristics of the RF cable that vary as a function of the
beam position, and localized attenuation/phase variations in the RF cable
characteristics.\\
\begin{figure}
\centerline{\includegraphics[width=28pc]{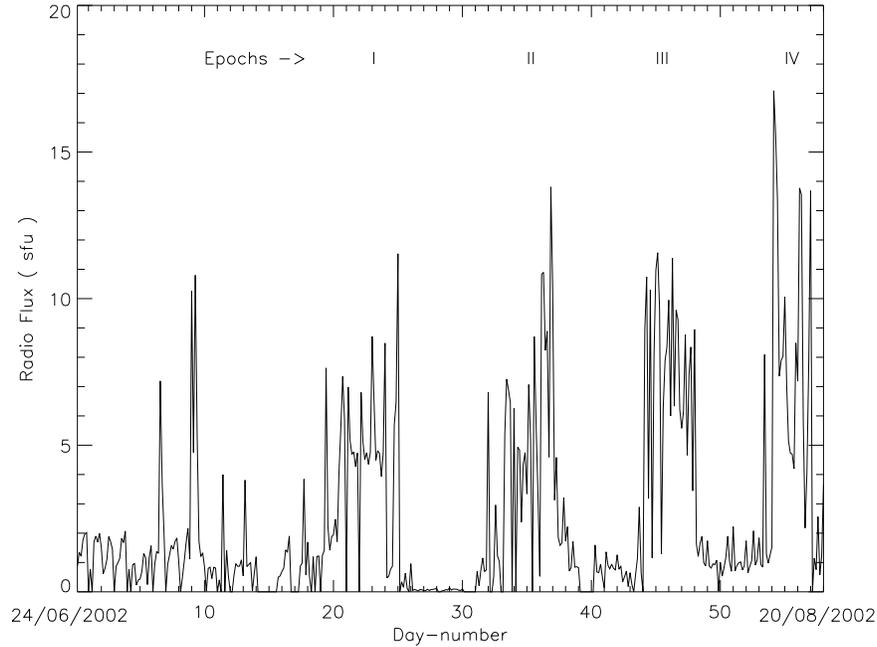}}
\caption{Solar radio flux as determined from tracking of the Sun consecutively
from 24th June 2002 to 20th August 2002; successive 7 data points
constitute a single day's observation. The four epochs of enhanced
radio emission are indicated by I, II, III, \& IV. The start- and end-dates
of the total observing schedule are indicated on the lower- left and right
corners, in the (dd/mm/yyyy) format.}\label{spf1}
\end{figure}
\\
The dwell-time ($t_{d}$) is defined as:
\begin{eqnarray}
t_{d} = t_{e} \times secant(\delta)~~~~~s.
\label{spe1}
\end{eqnarray}
where ($t_{e}$) is the period per degree of angular rotation in seconds
( $1^{o}~=~4$ minutes of time, for $\delta=0^{o}$).
Successive beams were positioned every $t_{d}$ minutes, as the radio source
changes in hour-angle, with the data being acquired concurrently.
This method was followed for about two months, during which period the
activity of the radio Sun underwent significant changes, as is evident
from Figure~\ref{spf1}. The plot depicts the calibrated solar radio flux
( obtained from averaging the data-samples about the transit point of
individual profiles ), as each data point, for a consecutive span of 58 days,
with each day's stake being 7 successive data sets ( from observations
made at the 7 beam positions everyday ). The data points were obtained
as a difference between the mean of 50 points in the total-power envelope,
symmetrically distributed about the transit point, and the median
of 50 data points of the profile-baseline.
In almost all cases where the radio flux attains a value of 0 sfu,
the absence of observational data
at the corresponding beam-position is being cited as the reason.\\
\\
The antenna-receiver performance characteristics were evaluated
extensively during the period, that comprised of extended quiet-Sun
epochs interspersed with solar activity, as a function of beam-position
in the end-to-end tracking window. Since the LPDA array for the
prototype tracking system has a group-beam size of $\sim 2.^{o}8$
( more than
the solar disk-size at 77.5 MHz ), a variation in the radio flux,
due to any particular solar radio activity, would have
attributes on a global scale, since the fine-structure components associated
with isolated regions on the solar disk are bound to be smudged-out by the
wider beam. Days of heightened solar radio emission,
measured from the enhanced global radio flux, were
segregated for study of their periodicity and distribution characteristics.
\subsection{Complementary observations}
Type I noise storm exhibit an increased
tendency to occur above bipolar active sunspot groups, with specific
accent on the maximum area occupied by a bipolar sunspot or the
spot-group,
and the active region complexity as regards its magnetic topology
( \cite{elg77}~and references therein ).
Studies performed by \cite{ddh57}~and~\cite{dun73}
on such a correlation, suggest values of
area associated with a sunspot group as low as a
100 millionth of that of the solar disk, as the minimum requirement.\\
\\
According to \cite{bru83}, all type I noise storms
observed during the Skylab period were caused by changes in the coronal
magnetic field structure,~and all coronal magnetic field changes observed
on the disk were correlated with newly emerging flux;
the storm source was found to
lie in the closed loop of strong magnetic fields,
just above the associated active region.
Hence it is clear that the existence of a sunspot
or a group of sunspots is a necessary condition for the generation of
noise storms.\\
\begin{figure}
\centerline{\includegraphics[width=28pc]{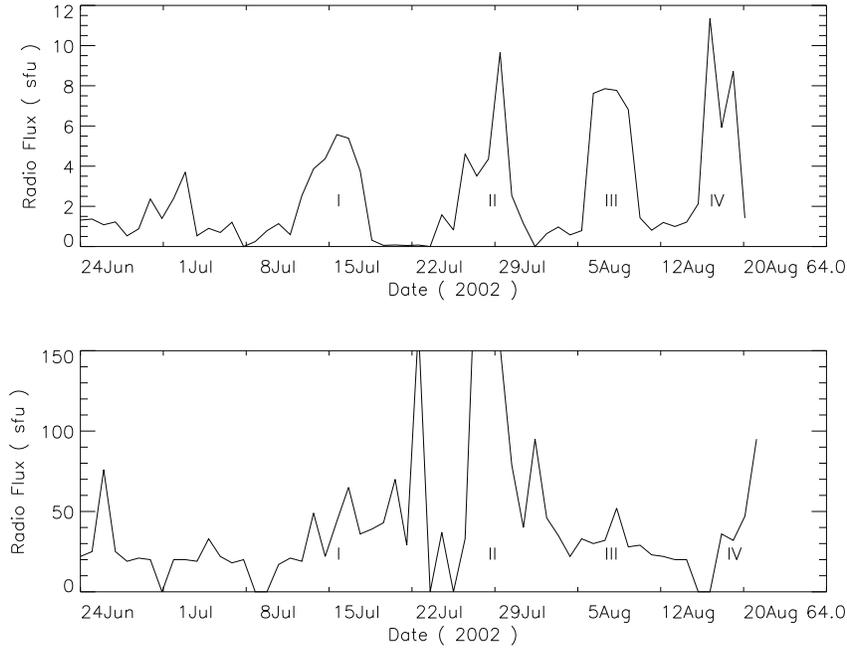}}
\caption{Daily mean total-Sun radio flux at 77.5 MHz ( GPTS : upper plot ),
and 245 MHz ( Sagamore Hill : lower plot )
from $24^{th}$ June to $20^{th}$ August, 2002.
The four epochs of global flux enhancement are denoted by I, II, III, \& IV.
Along the x axis,
spacing between adjacent dates constitutes a span of one week ( 7 days ).
\label{spf7}}
\end{figure}
\\
The noise storm radiation has been known to attain a sharp maximum
near the Central Meridian Passage ( CMP ) of a large sunspot group,
thereby revealing the sharply defined directivity characteristics.
The radiation has its origins within an area on the Sun's disk, that
is of the same order in size and angular position-coordinates to that
of the associated spot-group (~\cite{wld50} \& references therein ). Full-disk
$H{\alpha}$ images, from observations made at the BBSO,
reveal the presence of large and complex ARs
underlying the regions of enhanced noise storm radiation; the
temporal correlation between the CMP of the former with the
emission of the latter is the most noteworthy.\\
\\
In order to further illustrate the temporal association,
data from the GPTS, averaged over seven successive beam-positions
( daily mean flux ), is depicted alongside the daily mean
total solar radio flux measurements obtained at 245 MHz ( adjusted to 1 A.U.)
by the Sagamore Hill Observatory, Massachusettes, USA and published as
"Daily Solar Indices" in the Solar Geophysical Data ( SGD - prompt reports ),
for the same period.
Taking due cognizance of the missing data points in the GPTS data, and the
differing coronal processes at varying plasma levels ( that correspond to
77.5 MHz and 245 MHz ), the temporal correlation between the two sets
of data is very evident over the period considered for study.
A quantitative analysis for correlation,
with reference to the two plots in Figure~2,
yielded values for their coefficients
of 0.78, 0.8, 0.96, \& 0.56 for the four periods
( viz., I, II, III, \& IV ) of enhanced radio-flux
emission - indicative of a fairly reasonable correlation.
\section{Data Analysis}
\begin{figure}
\begin{center}
\epsfxsize=20cm
\epsfysize=35cm
\rotatebox{90}{\epsfbox[100 -300 700 600]{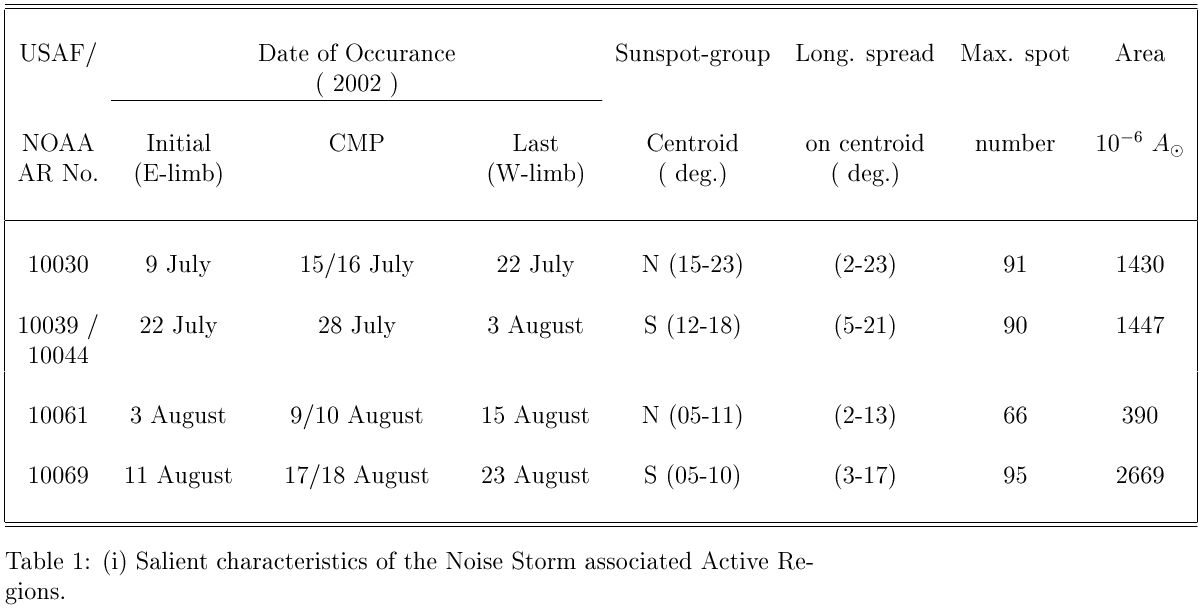}}
\end{center}
\end{figure}
\begin{table*}
\begin{center}
\begin{tabular}{|crclcc|}
\hline
& & & & &\\
&~& NRH - 164 MHz~& & Radio & Flux\\
& & & & &\\
\cline{2-4}\cline{5-6}
& & & & &\\
Date of &~& Heliographic~& & NRH & GPTS \\
CMP & & Positions & & &\\
\cline{2-4}
( 2002 ) & ( E - W ) & ( Mean Values ) & ( N - S ) & &\\
\cline{2-4}
& ( deg.) & & ( deg.) & ( sfu ) & ( sfu )\\
\hline
& & & & &\\
15/16 July & 25~W & & 21~N & 20-300 & 6.65\\
& & & & &\\
28 July & 13~W & & 18~S & $>$~300 & 7.4\\
& & & & &\\
9/10 August & 5~E & & 7~N & 100-300 & 11\\
& & & & &\\
17/18 August & 28~W & & 31~S & $\sim$~300 & 10.76\\
& & & & &\\
\hline
\end{tabular}
\caption[]{(ii)~~Salient characteristics of the Noise Storm associated
Active Regions, observed on the dates of CMP, with the NRH and GPTS.}
\label{spt2}
\end{center}
\end{table*}
\begin{table*}
\begin{center}
\begin{tabular}{|crcl|}
\hline
& & &\\
Date of & & Observing Time ( U.T.) &\\
CMP & & &\\
\cline{2-4}
& & &\\
( 2002 ) & NRH ( 164 MHz ) & & GPTS ( 77.5 MHz ) \\
& & & ( h.a.=$27^{o}$~W ) \\
& & &\\
\hline
& & &\\
16 July & 09:41:58 & & 09:47.67\\
& & &\\
28 July & 09:41:53  & & 09:48.33\\
& & &\\
9 August & 09:40:49 & & 09:43.44\\
& & &\\
18 August & 09:39:47 & & 09:40:00\\
& & &\\
\hline
\end{tabular}
\caption[]{Temporal coincidence in the NRH and GPTS observing times.}
\label{spt3}
\end{center}
\end{table*}
\begin{figure}
\tabcapfont
\centerline{%
\begin{tabular}
{c@{\hspace{0.3pc}}c}
\includegraphics[width=2.4in]{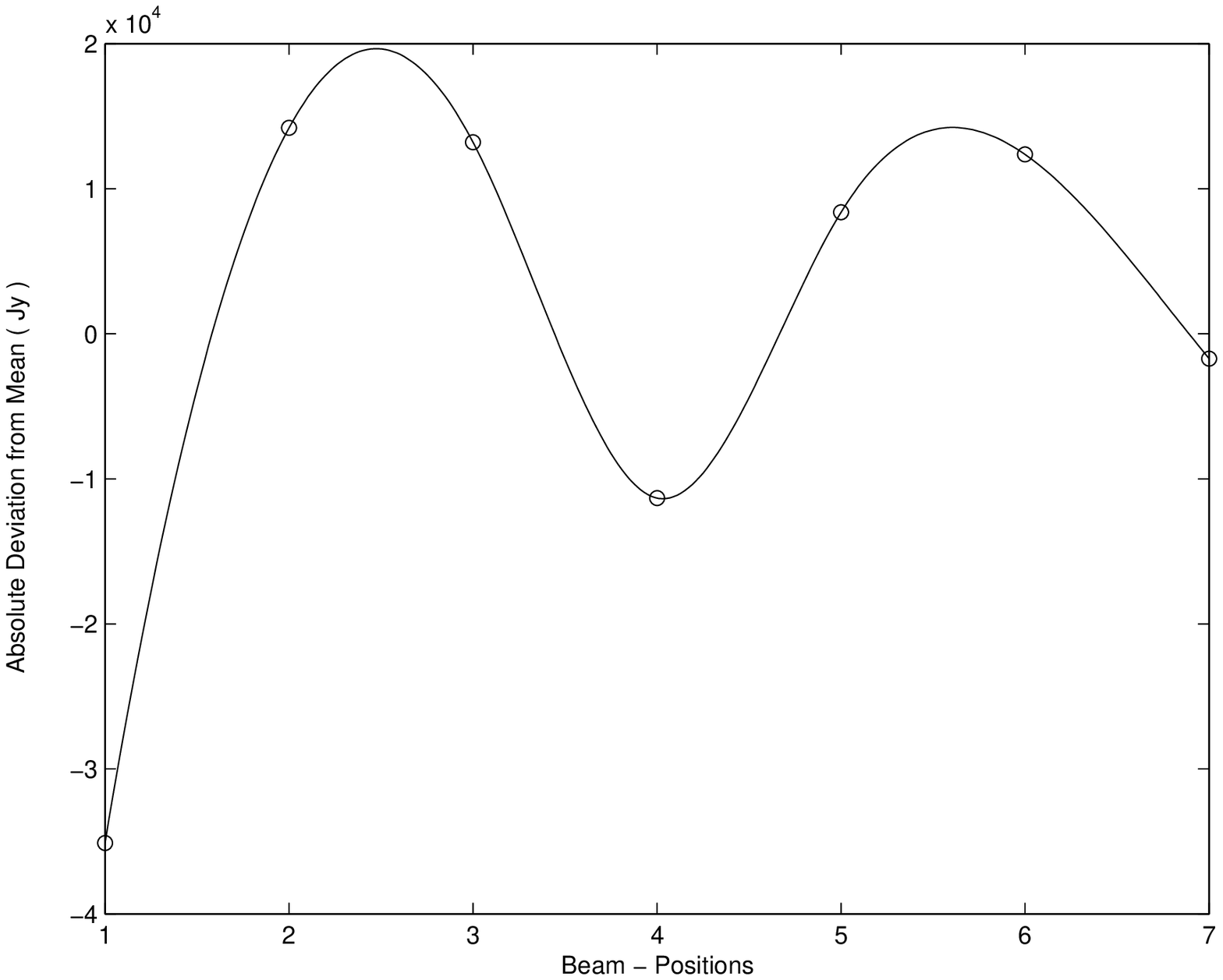} &
\includegraphics[width=2.4in]{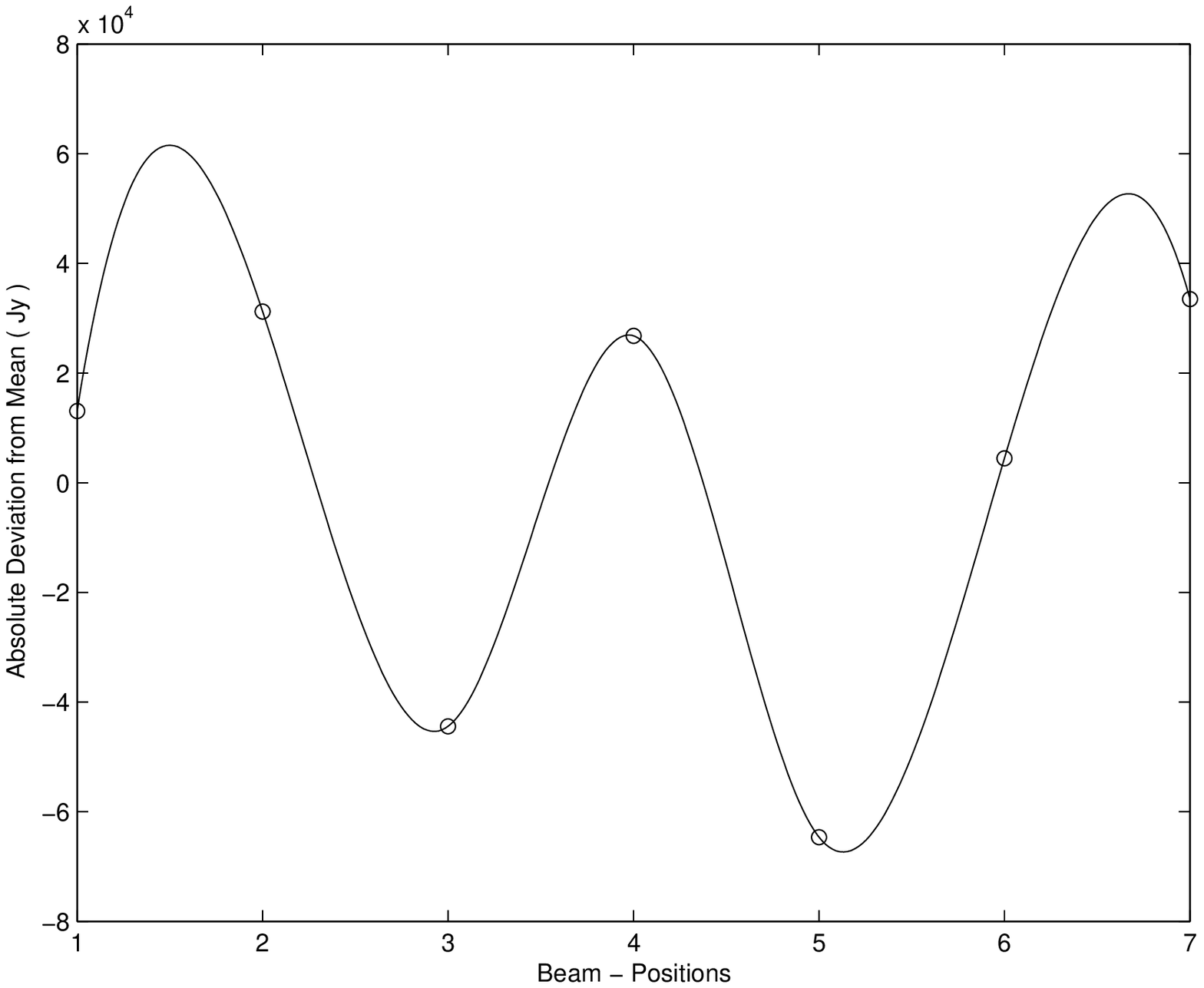} \\
\includegraphics[width=2.4in]{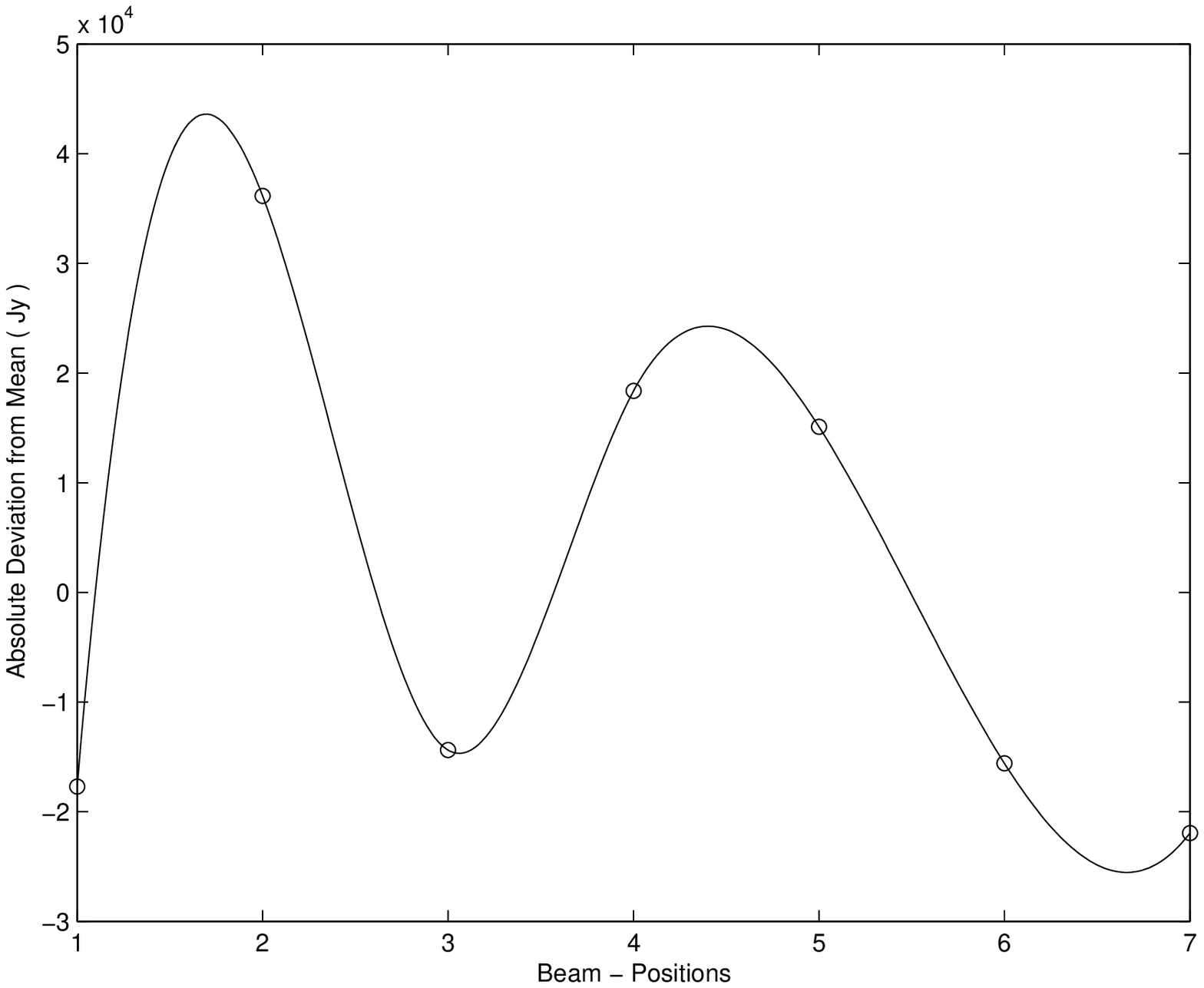} &
\includegraphics[width=2.4in]{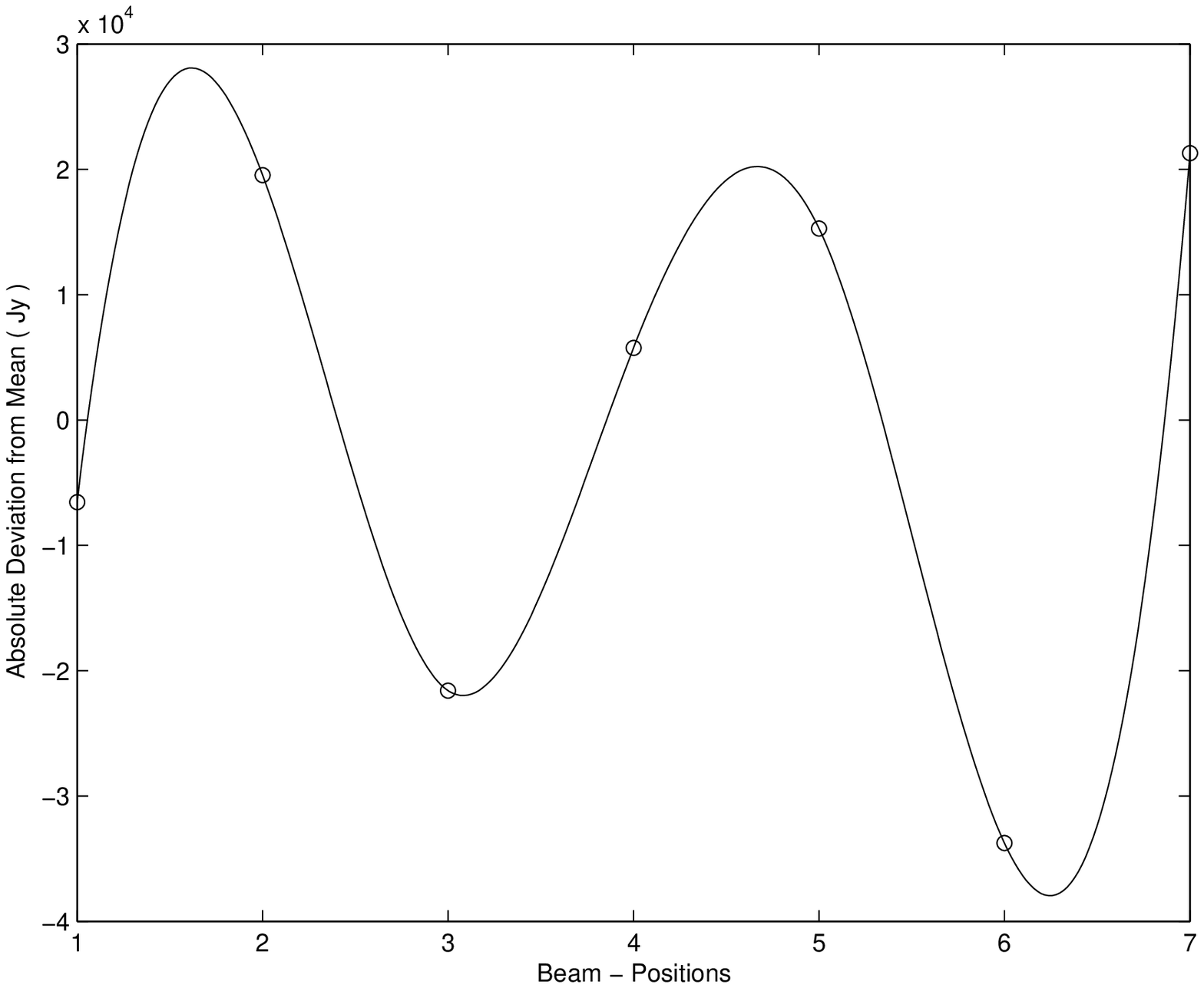} \\
\end{tabular}}
\caption{Scatter-plots of absolute deviation from mean solar radio flux
( in janskys ) with beam-position of GPTS on days of activity ( 28 July, and 7,
9~\&~10 August 2002 ); the solid line is a cubic-spline interpolation
fit to the data points.}\label{spf5}
\end{figure}
\begin{figure}
\centerline{\includegraphics[width=28pc]{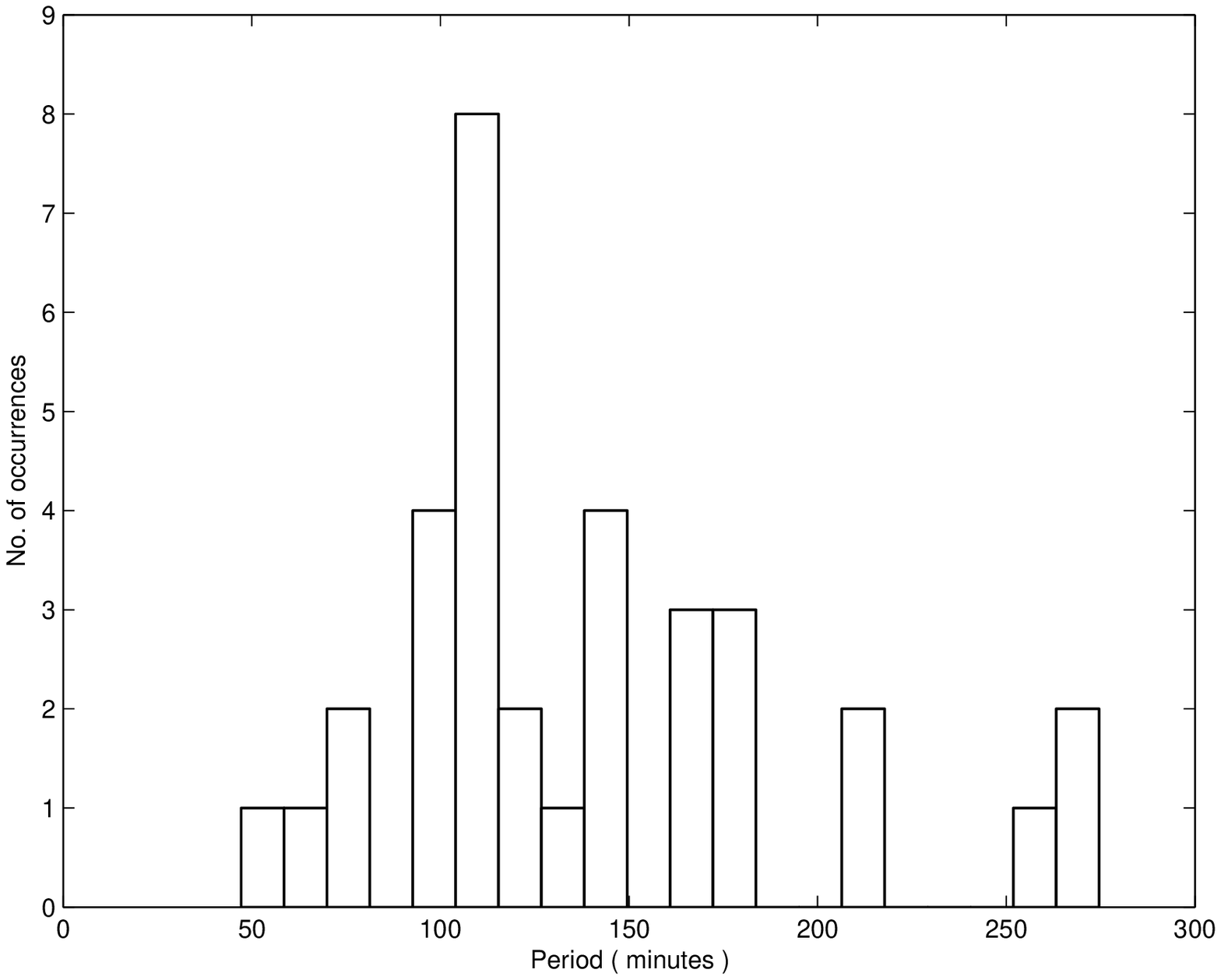}}
\caption{Distribution of the day-to-day absolute deviation from mean
value, with time in minutes, for the radio flux of noise storms at 77.5 MHz.
\label{spf6}}
\end{figure}
Table~I.(i) lists-out the salient characteristics of the large ARs
observed on the full disk $H{\alpha}$ images of BBSO.
The first column gives the USAF/NOAA active region
number, the II set of columns stand for the dates of occurrence ( during the
period of the noise storms considered ) of the ARs ( their initial appearance
on the East limb, cross-over at the Central Meridian and the last appearance
on the West limb ). The III \& IV columns give the coordinates of the
AR-centroid and the longitudinal extent about the centroid in degrees,
during the entire span in days, elapsed since their appearance and traversal
of the solar disk, as mentioned by the extremity dates of column II.
The data add greater significance to the earlier works cited in \S 3.1,
regarding the extent and
location of the noise storms vis-a-vis that of the underlying ARs, especially
since the coronal structures are magnetically confined to and rigidly towed along
by the underlying photospheric features~\cite{asc94}.
Column V \& VI list the maximum number of sunspots and area attained in the ARs.
The area comprising each of the sunspot groups is seen to be far
in excess of the minimum criterion ( which is a 100 millionth of that
of the projected visible solar hemispherical disk ) for the underlying
active region to be closely attributed to metric type I noise storm radiation.
The number density of the sunspots and the area occupied on the
solar disk are indicative of the complexity of the AR;
this, and the dates of CMP of storm centers at the
coordinates indicated in Table~\ref{spt2}.(ii), strongly suggest a
spatio-temporal association of the ARs with the global flux
enhancements due to the noise storms as shown in Figure~\ref{spf1}.\\
\\
During the entire observing schedule spanning 58 days ( of $\sim 4$ hrs
each day ), periods of enhanced global solar radio emission were
detected over several days. They were grouped into four prominent
phases of enhanced solar radio flux, corresponding to prolonged
noise storm activity at the coordinates shown in Table~I.(ii), from
Nan\c cay Radioheliograph observations at 164 MHz. The table also
gives details on the dates of CMP of the associated ARs underlying the
metric noise storm sources, their heliographic positions
and calibrated peak flux estimates
in sfu, for the NRH and GPTS data.\\
\\
A unique temporal coincidence exists for the observations made by the
NRH at 164 MHz, with that of the last beam-position
( hour angle ( h.a.) = $27^{o}$~W ) of the GPTS at 77.5 MHz
( Table~\ref{spt3} ), thereby providing a direct means of spatially
( and by virtue of their near-similar observing times, temporally )
associating the radio flux enhancement, observed in the GPTS data,
with the particular event recorded as two dimensional imaging
information by the NRH.
As is discernible from Table~I.(ii), the flux values
detected by either of the instruments
allude to the noise storm events of 16th \& 28th
of July and 9th \& 18th of August, 2002.
Temporal coincidences with spectral observation
on type I noise storms at metric wavelengths, made by Potsdam,
Izmiran \& Culgoora observatories,
and listed in the Solar Geophysical Data : Prompt Reports~-~
\cite{sgdp2},~\cite{sgdp3}~\&~\cite{sgdp4},
corroborate the observations made with the tracking system during this period
of solar activity.\\
\\
With the threshold level, for classifying a particular data point as
being an instance of noise storm activity,
against a flux-poor background, set at 5 sfu, the total number
of days of enhanced noise storms was found to be 17, or a total of 119 data points.
On each of those days,
the absolute deviation from mean flux ( chosen to be the mean value
for radio flux about transit, among the seven beam position on that day )
was plotted as a scatter-plot,
as in Figure~\ref{spf5}. Curve-fitting, by cubic-spline interpolation performed
on the data points, reveals 32 cases of quasi-half-periods for absolute
deviation from the mean values.\\
\section{Discussion \& Conclusion}
The general consensus on the radiation mechanism of noise storms is
one of plasma emission. New emergence of magnetic flux, and the
changes in the topography of the magnetic field arising out of the
reconnection~\cite{bewen81} of ruptured, preexisting flux lines
with the newly sprouting flux,
about the sites
of bipolar sunspot activity, leads to
the formation of recurrent shock waves~\cite{spbeh81},~\cite{wen81}.
The electrons of the ambient plasma within the reconnection loop,
at the outer coronal layers,
become unstable as a result of this realignment of magnetic
lines of force, and generate the
Langmuir (L) waves or the Upper hybrid (UH) waves.
Such L-waves and UH-waves coalesce in turn with the
ion acoustic waves, at the sites of density inhomogeneities, to emit the type I
continuum noise storm radiation. On the other hand, bursts of type I
are produced when the UH-waves, produced by the electrons
trapped in magnetic flux loops, are scattered on the Lower hybrid (LH)
waves generated at their shock wavefront.\\
\\
The determination of quasi-periodicity is subject to a lower cut-off, defined
by the beam dwell-time. The adjacent
beam positions are $9^{o}$ apart, and the dwell-time varies as a
function of the apparent declination of the Sun, in a manner
defined by Equation~(\ref{spe1}),
where $t_{e}$ is 36 minutes, and ($\delta$)
varied from $13.^{o}2$~N to $23^{o}$~N. Hence, $t_{d}$ is the minimum value
for quasi-periodicity that can be realised by the tracking system in its
current scheme of observation.
The quasi-periodic distribution, for the 32 cases of absolute deviation from mean,
is shown in the histogram of Figure~\ref{spf6}. The peak in the distribution
occurs at 110 minutes, which also is tantamount to the periodicity in global
solar radio flux variation at 77.5 MHz. In addition, more than half
($\approx 60 \% $)~of the
distribution occurs in the range of periodicities from 100-150 minutes.\\
\\
Regarding the spline interpolation method employed for determination
of the absolute deviation from mean ( shown in the four
sub-plots of Figure~\ref{spf5} ), it needs to be duly emphasised that
vast deviations, occuring on either side of the mean and at consecutive
beam position observations, are bound to yield quasi half-periodicity
values lesser than the minimum achieveable "cadence" between the two adjacent
observations on the same day - a case of so-called "super-resolution",
that needs to be cautiously approached, especially when the distribution
were to peak at those values. In Figure~\ref{spf6}, the first couple of
values in the vicinity of 50 min are the case in point, corresponding to the
first negative going half-waveforms on the upper-right and the lower-left
sub-plots of Figure~\ref{spf5}; they have values for quasi-periodicity less
than the $t_{d}$ defined by Equation~(\ref{spe1}) for the particular days.\\
\\
Quasi-periodic fluctuations in solar coronal emission, observed at the
operating frequencies of the antennas on-board the
Prognoz 1 high-apogee satellites~\cite{gps77} equipped with
kilometric radiation detectors,
have estimated periods ranging from 6 sec to 2 hours.
A specific quasi-periodicity value of 118 ($ \pm 20$)~minutes has been
quoted among many other values detected, for fluctuations in radio emission
by 10-15 dB from the mean solar flux. In the present study done at 77.5 MHz,
the quasi-periodic fluctuations have an absolute deviation from the
intra-day mean ranging from 1.5 to 4 sfu.\\
\\
The absolute deviation from mean flux, for each of the
17 days of enhanced activity, is taken as a measure of the intra-day
quasi-periodicity in solar radio flux, and found
to be 110 minutes, with the fluctuations in flux being $3 (\pm 1.5)$ sfu.
Positional information from the Nan\c cay RadioHeliograph data, and
features of the causative ARs of the underlying photospheric
disk from the full disk $H{\alpha}$ images of the BBSO,
along with the radio spectral data published in the
SGD reports, lead to the conclusion that heightened flux emission,
with global ramifications, are a result of type I noise storms.\\
\\
The occurrence of the noise storm source regions, and their
contribution to enhanced radio emission on a global scale, has been
corroborated from complementary evidence based on active region
data obtained from the BBSO full disk $H{\alpha}$ images, the NRH's
164 MHz imaging data on noise storms and the metric noise storm
spectral observations as reported in the SGD reports. Quasi-periodic
pulsations in global solar radio flux were observed, and their origin
has been attributed to modulation of the plasma radiation by
magnetohydrodynamic ( MHD ) disturbances in the corona, at sites above
large ARs threaded by complex magnetic flux tubes.
MHD disturbances are generated
by the weak shocks associated with magnetic reconnection events,
as MHD oscillations or resonance of MHD
waves, and fluctuations ( pulsations ) in the
plasma radio emission ensue at the source region for noise storms.
MHD waves on all scales, ranging in wavelength from
the coronal loop-size
(fraction of $R_{\odot}$) down to the gyroradii (a few meters) of coronal
ions, are believed to play a key role in the transport of mechanical energy,
from the denser regions of the chromosphere to the Sun's corona and further
as a steady stream called the solar wind; by means of dissipation
of the wave energy, the corona is heated and sustained at elevated temperatures.
The emissivity of trapped epithermal particles in coronal magnetic arches
is modulated by a propagating MHD wave~\cite{asc87}.
In the case of noise storms, the plasma waves get converted to
transverse electromagnetic ( TEM ) waves at sites of trapped density
inhomogeneities and magnetic reconnections.
Theoretical interpretation of the oscillatory phenomena, observed in the
outer solar corona, based on the MHD wave theory, along with supportive
information, regarding the plasma parameters, from investigation of
high spatial and temporal imaging data on the coronal
magnetic structures above the active region complexes, would significantly
demystify the underlying mechanisms involved in the plasma
dynamics of the outer corona, coronal-seismology and coronal-heating.
\acknowledgements
We thank the scientific and technical staff of the Gauribidanur
Radio Observatory, and the Nan\c cay Radioheliograph team
at the Unite Scientifique de Nan\c cay
( Station de RadioAstronomie de Nan\c cay,  18330 Nan\c cay,
France ) of the Observatoire de Paris for use of their
2D radio maps.
We also thank the referee for insightful comments which improved
the content and presentation of this paper.
The Solar Geophysical Data Reports are
published by the National Geophysical Data Center, NOAA, 325
Broadway, Boulder, Colorado, 80305-3328 USA. The $H{\alpha}$
images were obtained from the FTP Data Archive and World Wide Web pages
of the Big Bear Solar Observatory/New Jersey Institute of Technology,
40386 North Shore Lane, Big Bear City, CA  92314~USA.

\theendnotes

\printindex
\label{lastpage}
\end{article}
\end{document}